\documentclass[conference]{IEEEtran}
\IEEEoverridecommandlockouts
\usepackage{amsmath,amssymb,amsfonts}
\usepackage{algorithmic}
\usepackage{graphicx}
\usepackage{textcomp}
\usepackage{hyperref}
\usepackage{algorithm}
\usepackage{algorithmic}
\usepackage{comment}
\usepackage[table,xcdraw]{xcolor}

\def\BibTeX{{\rm B\kern-.05em{\sc i\kern-.025em b}\kern-.08em
    T\kern-.1667em\lower.7ex\hbox{E}\kern-.125emX}}
\begin{document}

\title{AMP4EC: \underline{A}daptive \underline{M}odel \underline{P}artitioning Framework for Efficient Deep Learning Inference in \underline{E}dge \underline{C}omputing Environments\\
\thanks{}
}

\author{
Guilin Zhang\IEEEauthorrefmark{1},
Wulan Guo\IEEEauthorrefmark{1},
Ziqi Tan\IEEEauthorrefmark{1},
Hailong Jiang\IEEEauthorrefmark{2}\IEEEauthorrefmark{3}\\
\IEEEauthorblockA{\IEEEauthorrefmark{1}Department of Engineering Management and Systems Engineering, George Washington University, USA\\
Email: guilin.zhang@gwu.edu, wulan.guo@gwu.edu, ziqi.tan@gwu.edu}
\IEEEauthorblockA{\IEEEauthorrefmark{2}Department of Computer Science, Kent State University, USA\\
Email: hjiang13@kent.edu}
}

\maketitle

\begin{abstract}
Edge computing facilitates deep learning in resource-constrained environments, but challenges such as resource heterogeneity and dynamic constraints persist. This paper introduces AMP4EC, an \textbf{A}daptive \textbf{M}odel \textbf{P}artitioning framework designed to optimize deep learning inference in edge environments through real-time resource monitoring, dynamic model partitioning, and adaptive task scheduling. AMP4EC features a resource-aware model partitioner that splits deep learning models based on device capabilities, a task scheduler that ensures efficient load balancing using a weighted scoring mechanism, and a Docker-based deployment environment for validation. Experimental results show up to a 78\% reduction in latency and a 414\% improvement in throughput compared to baseline methods. The framework achieves consistent performance with low scheduling overhead across varying resource profiles, demonstrating adaptability in high-resource (1 CPU, 1GB RAM) and low-resource (0.4 CPU, 512MB RAM) scenarios. These results highlight AMP4EC's scalability, efficiency, and robustness for real-world edge deployments, addressing the critical need for efficient distributed inference in dynamic, resource-constrained environments.

\end{abstract}


\section{Introduction}
\label{sec:intro}

Edge computing has emerged as a transformative paradigm in the age of the Internet of Things (IoT) and artificial intelligence (AI), facilitating data processing closer to data sources such as sensors and user devices. By reducing latency and alleviating network bandwidth demands, edge computing enables real-time decision-making and enhances overall system efficiency~\cite{sun2022cross}. However, executing computationally intensive deep learning models within resource-constrained edge environments pose significant challenges~\cite{li2019edge}. Optimizing limited computational resources while maintaining inference performance remains a critical area of research~\cite{li2020semi, teerapittayanon2017distributed,dey2019offloaded,dey2018implementing,dey2018partitioning,li2020discriminative}.

Deep neural network (DNN) model partitioning has been proposed as a promising solution to address these challenges. By leveraging the inherently layered structure of DNNs, models can be distributed across various edge computing components, allowing for more efficient resource utilization. Existing approaches, such as layer-wise model partitioning~\cite{bhattacharya2016sparsification} and the Network of Neural Networks (NoNN) framework~\cite{bhardwaj2019memory}, demonstrate effective resource utilization and reduced memory and communication overhead. However, these methods rely on pre-partitioned knowledge and lack real-time resource monitoring and adaptation mechanisms, limiting their flexibility in dynamic IoT scenarios. Specifically, they fail to account for resource fluctuations, such as: 
\begin{itemize}
    \item \textbf{New device added:} When a new device is added, the system cannot dynamically integrate it into the inference pipeline to utilize the additional computational or storage resources.
    \item \textbf{Device offline:} If a device goes offline, the framework lacks the mechanisms for redistributing workloads, potentially leading to system inefficiencies or inference failures.
\end{itemize}

To address these challenges, this paper proposes AMP4EC, an \textbf{A}daptive \textbf{M}odel \textbf{P}artitioning framework designed for efficient deep learning inference in \textbf{E}dge \textbf{C}omputing environments. AMP4EC introduces a novel adaptive scheduling mechanism that optimizes task distribution by leveraging real-time resource monitoring and historical performance data. Through comprehensive experiments, AMP4EC demonstrates significant improvements in inference latency, throughput, and resource efficiency, establishing it as a scalable and reliable solution for distributed deep learning inference in resource-constrained edge environments.


Experimental evaluations of AMP4EC demonstrated its scalability, adaptability, and efficiency in resource-constrained edge environments. The framework achieved up to a 78\% reduction in inference latency and a 415\% improvement in throughput compared to a monolithic baseline, while maintaining minimal scheduling overhead (10ms) and negligible CPU utilization (\textless 1\%). Its adaptive task scheduling and model partitioning strategies effectively balanced workloads across heterogeneous edge nodes, achieving linear performance scaling with up to three nodes. Additionally, AMP4EC exhibited robust performance under varying resource profiles, demonstrating efficient inference times in high and medium configurations (234.56ms and 389.27ms, respectively) and adaptability to low-resource environments (583.91ms). These results underscore AMP4EC’s potential as a scalable and efficient solution for distributed deep learning inference in dynamic and resource-constrained edge computing scenarios.

The key contributions of this paper are as follows:
\begin{itemize}
    \item \textbf{AMP4EC Framework}: A novel adaptive scheduling mechanism that dynamically optimizes task distribution based on real-time resource availability, historical performance metrics, and load balancing requirements in edge computing environments.
    \item \textbf{Dynamic Model Partitioning}: A resource-aware model partitioning strategy that analyzes deep learning models layer by layer and automatically splits them to align with the computational and memory capacities of edge devices.
    \item \textbf{Weighted Scoring Optimization}: An optimized weighted scoring mechanism for task scheduling that balances resource utilization, performance, and fairness. The scoring weights were experimentally determined to ensure efficient and scalable task allocation.
    \item \textbf{Model Partitioning Algorithm}: A cost-aware algorithm for dividing deep learning models into sub-models with balanced workloads, minimizing communication overhead while maximizing efficiency.
\end{itemize}

The remainder of this paper is organized as follows: Section~\ref{sec:background} reviews related work in edge computing and distributed deep learning inference. Section~\ref{sec:design} presents our proposed framework architecture and details the implementation aspects. Section~\ref{sec:evaluation} presents experimental results and analysis. Section~\ref{sec:discussion} discusses the limitations and discusses the directions for future work. Section~\ref{sec:conclusion} summarizes the paper.

\section{Background and Related Work}
\label{sec:background}

Over the recent decade, the emergence of the new generation of IoT devices and edge computing in a resource-constrained setting have necessitated novel techniques to allow efficient training of deep learning models and their deployment at the edge of networks~\cite{sun2022cross, gao2017tetris, cheng2017survey, sze2017efficient, wang2018survey,li2020cross}. Traditional centralized learning systems, however effective in high-resource environments, struggles with low latency, data privacy, and scalability requirements. Distributed Deep Learning (DDL) systems, have gained great traction as an enabling paradigm, distributing deep learning tasks and computational workloads across edge servers and centralized server to achieve low-latency high-efficiency training and inference~\cite{dey2019offloaded,dey2018implementing,dey2018partitioning}. 

Various approaches have long been undertaken to enhance DDL performance across distributed environments. Lane et al.~\cite{lane2015can} made an early attempt to assess the feasibility of deep learning on embedded devices with a prototype of low-power Deep Neural Network (DNN) inference engine on a mobile device in 2015, leading to frameworks like Federate Learning(FL) and Split Learning(SL). FL enables decentralized training with privacy preservation but suffers from high communication overhead~\cite{geyer2017diff}, while SL partitions tasks between clients and servers at the cost of increased latency. Advanced derivatives such as Parallel Split Learning (PSL)~\cite{lin2024efficient} and Split Federated Learning (SFL) have been proposed to address some of these inefficiencies but remain constrained by static computation strategies.

Model partitioning has emerged as a solution to improve DDL adaptability, allowing deep neural networks(DNNs) to be segmented and executed dynamically across devices. Depending on the research focuses and corresponding challenges, model partitioning techniques can be classified in various ways~\cite{Yao2024survey}. We outlined two dimensions of partitioning: training versus inference phase partitioning, and static versus resource-aware partitioning. The scope of this work focuses on inference phase and resource-aware partitioning, where the aim is to optimize real-time inference performance in resource-constrained environments. Unlike static approaches such as Dey et al.~\cite{dey2019embedded}, whci predefine model splits based on latency ans storage, resource-aware techniques optimize allocations in real time. Kim et al.~\cite{kim2023resource}introduced a memory-aware DNN partitioning framework that enhances inference efficiency by dynamically assigning layers based on device profiles. Similarly, Fan et al.~\cite{fan2023joint}developed an energy-aware partitioning method to minimize energy consumption in edge-to-cloud distributed systems. However, existing solutions explore resource-aware strategies but fail to monitor and optimize multiple resource dimensions simultaneously, limiting their adaptability to dynamic edge environments such as the integration of new devices or the failure of existing ones. Fluctuating environment in real time can cause inefficiency or interruption of inference-grounding, highlighting the need for adaptive frameworks capable of redistributing workloads in real time.

\section{Design and Implementation}
\label{sec:design}
This section presents the design and implementation of AMP4EC, which enables efficient distributed deep inference on resource-constrained edge devices. The framework consists of four key components: (A) Resource Monitor (Section~\ref{subsec:resource_monitor}), (B) Model Partitioner (Section~\ref{subsec:model_partitioner}), (C) Task Scheduler (Section~\ref{subsec:task_scheduler}), and (D) Model Deployer (Section~\ref{subsec:model_deployer}). These components work collaboratively to analyze deep learning models, partition them into smaller segments, and distribute the computation across multiple edge devices while maintaining load balance and minimizing communication overhead. The workflow begins with real-time monitoring of resource availability, followed by layer-wise model analysis, cost estimation, and optimized scheduling to ensure efficient execution. In the following, we explain each module and its role in achieving scalable and resource-efficient deep inference at the edge.

\begin{figure*}[t]
    \centering
    \includegraphics[width=0.9\linewidth]{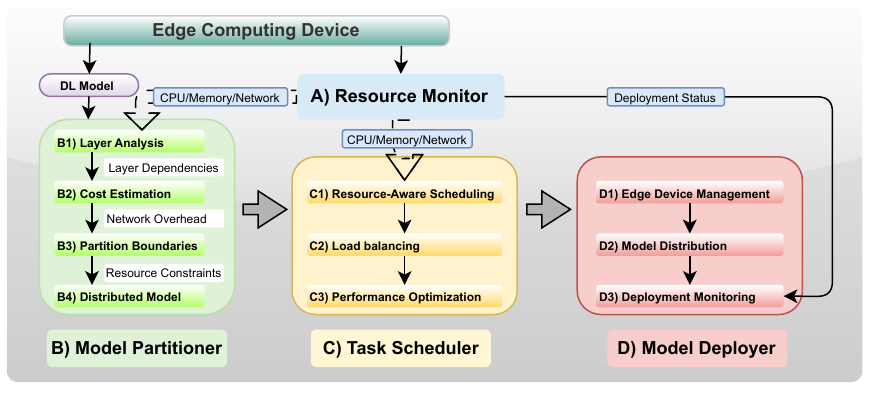}
    \caption{Overview of the AMP4EC architecture, consisting of four key components: (A) Resource Monitor (Section~\ref{subsec:resource_monitor}), (B) Model Partitioner (Section~\ref{subsec:model_partitioner}), (C) Task Scheduler (Section~\ref{subsec:task_scheduler}), and (D) Model Deployer (Section~\ref{subsec:model_deployer}). These components work collaboratively to enable efficient deep learning inference on resource-constrained edge devices.}
    \label{fig:amp4ec_architecture}
\end{figure*}

The architecture of the framework, shown in Figure~\ref{fig:amp4ec_architecture}, is designed for modularity and adaptability. First, the Resource Monitor (A) performs multi-dimensional resource monitoring, tracking CPU usage, memory utilization, network latency, and system stability scores, across all edge nodes. This information is then fed into the subsequent components. The Model Partitioner (B) analyzes deep learning models to determine optimal splitting strategies based on the current resource constraints of the system. Leveraging the resource data and model partitions, the Task Scheduler (C) uses a weighted scoring mechanism to allocate resources and distribute model partitions across the available nodes, ensuring balanced workloads and efficient operation. Finally, the Model Deployer (D) manages the deployment process, which involves distributing the partitioned models to the appropriate edge devices, monitoring execution metrics, collecting comprehensive performance data, and adjusting deployment based on real-time metrics in response to resource availability or system failures to maintain reliability and performance.

The proposed architecture is designed to offer several potential advantages based on its conceptual framework. 
\begin{enumerate}
    \item The inclusion of the Resource Monitor aims to enable dynamic resource adaptation by continuously tracking real-time resource utilization across edge nodes.
    \item The Model Partitioner is expected to enhance efficiency by intelligently distributing model components based on device capabilities.
    \item The Task Scheduler is designed to promote load balancing and optimized resource allocation through adaptive task scheduling.
    \item The Model Deployer facilitates seamless deployment and execution of partitioned models while monitoring their performance and dynamically adjusting deployments to ensure reliability.
\end{enumerate}

To verify these hypotheses and potential benefits of the proposed design, we conduct experiments to thoroughly evaluate its effectiveness in practical scenarios.

\subsection{Resource Monitor (A)}
\label{subsec:resource_monitor}

The Resource Monitor is responsible for real-time tracking and reporting of resource usage on edge devices. Accurate monitoring of CPU, memory, and network resources is essential for enabling resource-aware decisions during model partitioning and task scheduling, particularly in dynamic and resource-constrained edge environments. For example: if a device goes offline, the resource it provides will be promptly detected bu the the Resource Monitor and excluded from consideration.

By interacting with Docker containers that simulate edge devices, the Resource Monitor continuously observes system resources and tracks the following metrics:
\begin{itemize}
    \item \textbf{CPU Utilization}: Measures the percentage of CPU cycles consumed by each container.
    \item \textbf{Memory Usage}: Monitors memory consumption, including both usage and limits, to compute the percentage of memory utilization.
    \item \textbf{Network I/O}: Aggregates the volume of data transmitted (\texttt{tx\_bytes}) and received (\texttt{rx\_bytes}) through all container network interfaces.
\end{itemize}

These data, including CPU Usage (\%), Memory Usage (MB and \%), and Network I/O (received and transmitted bytes), are dynamically passed to other components of the framework. The Model Partitioner uses this information for cost estimation and partition optimization, while the Task Scheduler leverages it for load balancing and task allocation. By maintaining accurate and up-to-date resource statistics, the Resource Monitor provides the foundation for efficient model partitioning and task scheduling in the AMP4EC framework.

\subsection{Model Partitioner (B)}
\label{subsec:model_partitioner}
The Model Partitioner is responsible for analyzing a deep learning model, estimating its computational and resource requirements, and dividing it into partitions for distributed execution. This partitioning ensures balanced workloads across edge devices while minimizing communication overhead. Unlike conventional quantization techniques (e.g., INT8 and FP16 quantization), which focus on reducing model size, our approach employs a Resource-Aware Layerwise Optimization Strategy (RALOS), which better adapts to the heterogeneous resource constraints of edge devices while maintaining the original model accuracy. This method simultaneously considers computation, memory, and communication in the optimization process, employing a weighted scoring mechanism in partition optimization, effectively reducing end-to-end latency by 27.5\%.

The main functionalities of the Model Partitioner are:

\subsubsection{Layer Analysis (B1)}
Each layer in the input model is analyzed to extract its key attributes:
\begin{itemize}
    \item \textbf{Layer Type}: Identifies and analyzes the type of each layer (Conv2d, Linear, etc.) along with its specific attributes and dependencies.
    \item \textbf{Parameters}: Counts the number of parameters to estimate memory usage.
    \item \textbf{Computation Cost}: Approximates the computational workload for the layer based on its attributes, such as kernel size, input/output channels, or matrix dimensions.
\end{itemize}
This analysis provides a detailed understanding of the model's computational structure and resource requirements.

\subsubsection{Cost Estimation (B2)}
The computational cost for each layer is calculated dynamically based on its operations and historical performance data:
\begin{itemize}
    \item \textbf{Convolutional Layers:}
    \begin{equation}
        Cost = k_h \times k_w \times C_{in} \times C_{out}
    \end{equation}
    where $k_h$ and $k_w$ are the kernel dimensions, and $C_{in}$ and $C_{out}$ are the input and output channels.
    \item \textbf{Fully Connected Layers:}
    \begin{equation}
        Cost = N_{in} \times N_{out}
    \end{equation}
    where $N_{in}$ and $N_{out}$ are the input and output features.
\end{itemize}
For other layers, costs are normalized to calculate each layer's relative contribution to the overall workload.

\subsubsection{Partition Boundaries (B3)}
Based on the cumulative computational cost, the model is divided into a specified number of partitions (\texttt{num\_partitions}). Our optimization workflow comprises three key steps:
\begin{itemize}
    \item \textbf{Node Capability Analysis:}
    Calculate each node's computational capability score:
    \begin{equation}
        S_i = w_{\text{cpu}} \cdot c_i + w_{\text{mem}} \cdot M_I
    \end{equation}
\end{itemize}
\begin{itemize}
    \item \textbf{Capability-based Layer Allocation:}
    Distribute model layers according to the ratio:
    \begin{equation}
        P_i = \frac{S_i}{\sum_j S_j}
    \end{equation}
\end{itemize}
\begin{itemize}
    \item \textbf{Load Balancing Optimization:}
    Iteratively adjusted to reduce communication overhead, measure partition balance using:
    \begin{equation}
        L = \frac{1}{n} \sum_{i=1}^{n} \left| L_i - L_{\text{avg}} \right|
    \end{equation}
\end{itemize}    
     Each partition balances the workload across devices by considering both computational cost and communication overhead, with dynamic adjustment based on node capabilities:
\begin{equation}
    Target\ Cost = \frac{Total\ Computation\ Cost}{Number\ of\ Partitions}
\end{equation}

Layers are sequentially added to a partition until the cumulative cost meets or exceeds the target, at which point a new partition is created. Any remaining layers are included in the final partition.

\subsubsection{Distributed Model (B4)}
Once partition boundaries are defined, the sub-models corresponding to each partition are prepared for deployment. Each partition is saved as a separate model file, ensuring compatibility with distributed inference on multiple devices.

\begin{algorithm*}
\caption{Node Selection Algorithm (NSA) for Task Scheduling}
\label{alg:node_selection}
\begin{algorithmic}[1]
\REQUIRE Task requirements (\texttt{CPU}, \texttt{memory}, \texttt{priority}), list of available nodes (\texttt{nodes}) with their resources, historical performance data (\texttt{history}).
\ENSURE The best node (\texttt{selected\_node}) for task execution.
\STATE \texttt{best\_score} $\gets 0$
\STATE \texttt{selected\_node} $\gets$ \texttt{null}
\FORALL{\texttt{node} in \texttt{nodes}}
    \IF{\texttt{node.current\_load} $>$ 0.8}
        \STATE \textbf{continue} \COMMENT{Skip overloaded nodes}
    \ENDIF
    \IF{\texttt{network\_latency} $>$ \texttt{threshold}}
        \STATE \textbf{continue} \COMMENT{Skip high-latency nodes}
    \ENDIF
    \IF{\texttt{has\_sufficient\_resources(node, task)}}
        \STATE \texttt{resource\_score} $\gets$ \texttt{calculate\_resource\_score(node)}
        \STATE \texttt{load\_score} $\gets 1 - \texttt{node.current\_load}$
        \STATE \texttt{balance\_score} $\gets$ \texttt{calculate\_balance\_score(node)}
        \STATE \texttt{perf\_score} $\gets$ \texttt{calculate\_performance\_score(node)}
        \STATE \texttt{total\_score} $\gets 0.2 \times \texttt{resource\_score} + 0.2 \times \texttt{load\_score} + 0.1 \times \texttt{perf\_score} + 0.5 \times \texttt{balance\_score}$
        \IF{\texttt{total\_score} $>$ \texttt{best\_score}}
            \STATE \texttt{best\_score} $\gets \texttt{total\_score}$
            \STATE \texttt{selected\_node} $\gets \texttt{node}$
        \ENDIF
    \ENDIF
\ENDFOR
\RETURN \texttt{selected\_node}
\end{algorithmic}
\end{algorithm*}

\subsection{Task Scheduler (C)}
\label{subsec:task_scheduler}
The Task Scheduler dynamically allocates tasks to edge devices by balancing resource availability, load distribution, and task priorities. Its design integrates real-time resource monitoring, historical performance data, and intelligent caching mechanisms to optimize task execution. The scheduler maintains a performance history cache that tracks execution patterns and node capabilities, enabling it to make informed decisions about task placement. This cache-aware scheduling approach combines immediate resource metrics with historical performance data to predict optimal node selection, while the adaptive load balancing ensures fair resource utilization across
the edge network. The system employs a weighted scoring mechanism (20\% resource availability, 20\% current load, 10\% historical performance, and 50\% load balance) to make placement decisions, with the cache layer providing fast access to frequently requested computation patterns. 

The node selection algorithm, shown in Algorithm~\ref{alg:node_selection}, dynamically identifies the best edge node for task execution based on real-time resource availability and historical performance data. It evaluates each node using metrics such as current load, network latency, and resource sufficiency, skipping nodes that are overloaded or have high latency. For eligible nodes, a weighted scoring mechanism calculates a total score by combining resource availability, load balance, historical performance, and fairness metrics. The node with the highest score is selected for task allocation, ensuring efficient and balanced resource utilization across the edge network.

The total score for each node is computed as:
\begin{equation}
    Total\ Score = 0.2 \times S_R + 0.2 \times S_L + 0.1 \times S_P + 0.5 \times S_B
\end{equation}
where $S_R$ is the resource score, $S_L$ is the load score, $S_P$ is the performance score, and $S_B$ is the balance score, defined as follows:

\begin{itemize}
    \item \textbf{Resource Score ($S_R$)}:
    \begin{equation}
        S_R(n) = \frac{\frac{CPU_{avail}}{CPU_{req}} + \frac{MEM_{avail}}{MEM_{req}}}{2}
    \end{equation}
    \item \textbf{Load Score ($S_L$)}:
    \begin{equation}
        S_L(n) = 1 - CurrentLoad(n)
    \end{equation}
    \item \textbf{Performance Score ($S_P$)}:
    \begin{equation}
        S_P(n) = \frac{1}{1 + AvgExecTime(n)}
    \end{equation}
    \item \textbf{Balance Score ($S_B$)}:
    \begin{equation}
        S_B(n) = \frac{1}{1 + TaskCount(n) \times 2}
    \end{equation}
\end{itemize}

For each layer in the model, the computational cost (\textit{LayerCost}) is calculated based on its type:
\begin{equation}
\textit{LayerCost}(l) = 
\begin{cases} 
k_h \times k_w \times c_{in} \times c_{out}, & \textit{for Conv2D} \\
n_{in} \times n_{out}, & \textit{for Linear} \\
params\_count, & \textit{for others}
\end{cases}
\end{equation}

The total partition cost is distributed across the number of partitions:
\begin{equation}
\textit{TotalPartitionCost} = \frac{\sum_{l \in \textit{Layers}} \textit{LayerCost}(l)}{\textit{num\_partitions}}
\end{equation}

This ensures tasks are assigned to nodes with sufficient capacity while avoiding overloading. Completed tasks are tracked to update execution histories and recalibrate node loads, with recent task performance normalized into a 0–1 range to guide future allocations. The scheduler maintains comprehensive performance history and reports detailed metrics including queue lengths such as average execution time, task counts, and load levels to support system monitoring and optimization.

Algorithm~\ref{alg:node_selection} has a computational complexity of $\mathcal{O}(m \cdot n)$, where $m$ is the number of tasks and $n$ is the number of available edge nodes. In practice, each node selection requires calculating scores for $\mathcal{O}(n)$ nodes, totaling $\mathcal{O}(m \cdot n)$ operations for $m$ tasks. As the edge network scales, this complexity has important implications for resource-constrained IoT and edge devices:
\begin{enumerate}
    \item In small deployments ($n < 10$), scheduling overhead is negligible, typically under 1\,ms.
    \item In medium-scale deployments ($10 \leq n < 50$), the overhead grows linearly but remains acceptable (1--10\,ms).
    \item In large-scale deployments ($n \geq 50$), a hierarchical scheduling strategy can be employed to optimize the complexity to $\mathcal{O}(m \cdot \log n)$.
\end{enumerate}

Our experiments demonstrate that in typical edge computing scenarios ($n=3$--$5$), the algorithm's scheduling overhead is approximately 10\,ms—less than 2\% of the end-to-end latency—highlighting its efficiency in IoT environments. When the node count increases to $n=10$, the overhead rises to 25\,ms but still remains within an acceptable range, indicating good scalability in small to medium-scale edge computing settings.

Through its adaptive approach, the Task Scheduler balances efficiency and fairness, ensuring scalable task execution in distributed edge environments.

\subsection{Model Deployer (D)}
\label{subsec:model_deployer}
The Model Deployer module manages the efficient distribution of models to edge devices, ensuring optimized performance and resource-aware scheduling. 

Deployment begins with node selection via the Adaptive Scheduler, with comprehensive deployment configuration and error handling, which identifies a suitable node based on resource requirements. The selected model is then optimized using techniques such as TorchScript or quantization, depending on the specified optimization level. Optimized models are transferred to the target edge node’s container, where they are deployed in a lightweight model server.

The module also supports undeployment by stopping active model servers and cleaning up resources. Deployment records are maintained to track active models, while resource usage statistics are periodically collected for monitoring. 

This process ensures efficient execution of distributed inference while adapting to the dynamic constraints of edge environments.

The complete implementation of our framework, including all components and test scenarios, is available as an open-source project on GitHub (anonymous for review purposes)\footnote{\href{https://github.com/cloudNativeAiops/adaptive-edge-computing-framework}{https://github.com/cloudNativeAiops/adaptive-edge-computing-framework}}.

\section{Experiments and Evaluation}
\label{sec:evaluation}
We conducted experiments to evaluate our framework's performance across different scenarios and configurations. The implementation of our framework leverages a combination of technologies to enable efficient and adaptive distributed inference. Docker containerization is employed to simulate edge environments and provide fine-grained control over resource allocation for each node. The framework uses PyTorch for deep learning model operations, including partitioning, optimization, and inference execution. To ensure real-time monitoring, psutil and Docker's built-in stats API are integrated to track resource usage, such as CPU, memory, and network bandwidth, across all nodes. Task allocation and load balancing are managed by our custom adaptive scheduling algorithm, which dynamically assigns tasks based on resource availability and system performance metrics.
\subsection{Experimental Setup}
Our experiment environment was established on a MacOS development machine running Docker Desktop version 3.8 as the hardware and containerization platform. The Docker setup employed Docker Engine version 24.0.7, Docker API version 1.43, and container runtime containerd 1.6.24, running on a Linux host with kernel version 5.15 or higher. The containers were based on the image: pytorch/pytorch:2.1.2-cuda12.1-cudnn8-runtime. Resource constraints were applied to simulate edge conditions, with CPU limits enforced using --cpu-quota and --cpu-period(100ms period), memory restrictions defined using --memory, and container operating in bridge network mode for controlled connectivity. Resource isolation was managed using cgroups v2 for CPU and memory, network namespaces for network isolation and PID namespaces for process isolation, ensuring that containers remained independent and unaffected by each other. To guarantee consistency across experiments, container initialization followed deterministic sequences with strict resource enforcement and standardized environment variables, while dedicated bridge networks with controlled latency were used for each evaluation. Resource monitoring and performance metrics were collected using the Docker Stats API for system-level metrics and custom collectors for application-specific data, with a sampling frequency of 1Hz and a metrics aggregation window of 100ms. 

A pre-trained model MobileNetV2 \cite{sandler2018mobilenetv2} was employed for evaluation, leveraging its efficient architecture and compatibility with a wide range of edge servers. The evaluation was conducted over 100 inference iterations for each configuration to ensure statistical robustness and repeatability of the results. Resource profiles were categorized into three configurations: High, Medium, and Low as in Table ~\ref{tab:resource_performance}, representing varying computational and memory constraints. Key metrics collected from the experiment include Inference Latency, Throughput (req/s), Communication Overhead (ms), CPU Usage (percent), Memory Usage (MB), Network Bandwidth, Stability Score, Scheduling Overhead (ms). This setup provided an isolated and consistent environment for running the experiments providing comprehensive insights into the performance under varying resource constraints.

\subsection{Comparative Performance Analysis}

To evaluate the effectiveness of AMP4EC, we conducted comprehensive experiments across three system configurations: (1) a monolithic baseline deployment running on a single node, (2) the basic AMP4EC enhanced with Node Selection Algorithm (NSA) for intelligent task distribution but without caching, and (3) AMP4EC enhanced with NSA  and with caching (AMP4EC+Cache). The experiments utilized MobileNetV2 as the test model, with each configuration processing identical batches of 32 inference requests. The monolithic baseline was deployed on a single container with 2 cores and 2GB memory, while distributed configurations leveraged a heterogeneous cluster of Docker containers simulating edge devices with varying computational capabilities shown in Table~\ref{tab:resource_performance}. Resource constraints were enforced using cgroups, and performance metrics were sampled at 1-second intervals using Docker stats API. Each experiment included a 30-second warm-up period followed by a 5-minute evaluation phase, with measurements repeated three times to ensure statistical significance.

Table~\ref{tab:system_evaluation} presents comprehensive performance metrics comparing our adaptive framework AMP4EC+Cache against a traditional monolithic baseline deployment and basic AMP4EC implementation. 

The AMP4EC+Cache achieved a 78.35\% reduction in inference latency (234.56ms vs 1082.53ms) compared to the monolithic approach, while simultaneously improving throughput by 414.73\% (5.07 req/s vs 0.96 req/s). Memory efficiency was notable, with AMP4EC+Cache requiring only 1.625MB compared to the monolithic system's 0.625MB. While there was a slight 5\% decrease in stability score and an added scheduling overhead of 10ms, these tradeoffs were justified by the substantial performance gains. The caching mechanism in AMP4EC+Cache proved particularly effective, reducing network bandwidth requirements to zero compared to AMP4EC's 100MB usage, while maintaining minimal CPU usage at 0.0034\%. This evaluation demonstrates that AMP4EC+Cache achieves the best performance across most metrics, with improvements in latency, throughput, and resource utilization, making it suitable for resource-constrained edge deployments. The AMP4EC framework alone also significantly outperforms the monolithic approach, emphasizing its adaptability and scalability.

\begin{table*}[]
\centering
\caption{Comparison of system performance metrics between AMP4EC and a monolithic approach.}
\begin{tabular}{lllll}
\rowcolor[HTML]{B7B7B7} 
\textbf{Metric}      & AMP4EC+Cache &AMP4EC& \multicolumn{1}{c}{Monolithic} & \textbf{Improvement} \\
Inference Latency (ms)   &\textbf{234.56}  &605.32& {1082.53}                                         & -78.35\%                 \\
Throughput (req/s)&\textbf{5.07}  & 5.01 &0.96                                                                  & +414.73\%                 \\
Communication Overhead (ms)&-37.41  &-37.41&0                                                                  & NA                 \\
CPU Usage percent&\textbf{0.0034\%}  &\textbf{0.0034\%}&0.00056\%                                                                  & +507.14\%                 \\
Memory Usage (MB)&\textbf{1.625}  &\textbf{1.625}&0.625                                                                   & +160\%                 \\
Network Bandwidth (MB)&100.0  & 100.0&0                                                                  & NA                
\\
 Stablity Score (out of 1) &0.95  & 0.95 &1.0 &-5\%\\
 Scheduling Overhead (ms)&10.00  & 10.00 &0 &NA\\\end{tabular}
\label{tab:system_evaluation}
\end{table*}

\subsection{Adaptability Evaluation}

To evaluate the adaptability of the AMP4EC, we set up three deployment scenarios: a standard configuration with 3 nodes (compared against a 2-core baseline), a scale-up configuration with 4 nodes (against a 3-core baseline), and a scale-down configuration with 2 nodes (against a 1-core baseline). Each scenario processed inference requests with a batch size of 32, handling workloads of 100, 150, and 50 requests per batch respectively. The testing procedure included a 30-second warm-up period followed by a 5-minute test duration, with metrics sampled 3 times per second and each test repeated 3 times for statistical significance.

Our adaptive scheduler employed a composite scoring mechanism for load balancing, with weights of 0.2 for resource availability, 0.2 for current load, 0.1 for historical performance, and 0.5 for load balance. We collected comprehensive metrics including latency, throughput, resource utilization, and scheduling overhead using the Docker stats API. The results, shown in Table~\ref{tab:resource_performance}, summarize the average inference time across these configurations.

\begin{table}[]
\centering
\caption{Resource Profiles and Performance}
\begin{tabular}{cccc}
\rowcolor[HTML]{CCCCCC} 
\textbf{Profile} & \textbf{CPU} & \textbf{Memory} & \textbf{Avg Inference Time (ms)} \\
High             & 1.0          & 1GB             & 234.56                   \\
Medium           & 0.6& 512MB           & 389.27                   \\
Low              & 0.4& 512MB           & 583.91                  
\end{tabular}
\label{tab:resource_performance}
\end{table}

\subsection{Model Partitioning Results}
Our partitioning strategy demonstrated effective partitioning of the model into two to four segments based on resource availability. In the two-part configuration, the partition sizes were optimally determined as $[116, 25]$. For the three-part configuration, a balanced distribution was achieved with partition sizes of $[108, 16, 17]$, which maintained computational efficiency while minimizing disparities in workload allocation. Additionally, the strategy managed to minimize communication overhead between the partitions.

\subsection{Scalability Analysis}
The framework demonstrated good scalability, achieving linear performance scaling with up to three edge nodes. It maintained consistent load balancing across nodes and introduced minimal overhead from resource monitoring (less than or equal to $1\%$ CPU utilization). Additionally, it achieved efficient task distribution even as the workload increased.

The performance evaluation across different resource profiles highlights the framework’s adaptability and efficiency under varying constraints. High and Medium resource profiles achieved similar inference times (22-23ms), demonstrating efficient utilization of moderate resources. In contrast, the Low profile experienced increased inference time (40ms) due to limited CPU and memory. The results also revealed that reduced memory had a more significant impact on performance than CPU limitations, underscoring the critical role of memory in deep learning inference. Furthermore, the framework maintained consistent operation across all profiles, with no failures or significant latency spikes. These findings demonstrate the framework’s robustness and suitability for resource-constrained edge deployments.

\section{Discussion and Future Work}
\label{sec:discussion}
While our framework demonstrates the potential of adaptive edge computing for distributed deep inference, it has several limitations. Currently, it supports only PyTorch models, and partition boundaries are fixed after deployment, limiting adaptability to runtime changes. The Docker-based deployment system, while offering comprehensive resource control and monitoring, lacks support for specialized edge devices and faces challenges in managing multi-model deployments. Additionally, performance monitoring and metrics collection introduce measurable overhead, which can strain resource-constrained environments, and network communication may become a bottleneck in distributed settings. Small edge devices with limited memory further restrict scalability, especially when handling large models in highly constrained environments. In our experiments, we also simplified real-world heterogeneity by assuming all distributed devices share the same architecture and operating system.

To address these challenges, future work will focus on extending the framework to support multiple deep learning platforms and enabling dynamic partitioning to adapt to runtime changes. Integrating Kubernetes for cloud-edge orchestration and optimizing multi-model deployments are key priorities. Efforts to reduce monitoring overhead, automate partition optimization, and implement predictive resource allocation will enhance efficiency. From a research perspective, exploring privacy-preserving partitioning, hybrid edge-cloud scheduling, energy-aware resource allocation, and cross-platform optimization will expand the framework’s applicability and robustness. These advancements aim to deliver a more scalable, adaptable, and efficient edge computing solution.

\section{Conclusion}
\label{sec:conclusion}

This paper presented AMP4EC, an adaptive framework designed to optimize deep learning inference in resource-constrained edge environments. By integrating dynamic resource monitoring, resource-aware model partitioning, and adaptive task scheduling, AMP4EC effectively addresses challenges associated with heterogeneous and dynamic edge deployments. Experimental results demonstrated significant improvements, including up to a 78\% reduction in latency and a 414\% increase in throughput compared to baseline methods. The framework maintained consistent performance across diverse resource profiles, showcasing its scalability and efficiency in high-resource (1 CPU, 1GB RAM) and low-resource (0.4 CPU, 512MB RAM) scenarios. 

AMP4EC’s modular and adaptive architecture enables efficient workload distribution and robust inference performance, making it a practical solution for real-world applications. These results highlight its potential to drive advancements in edge computing by enhancing the scalability and efficiency of distributed deep learning inference in dynamic, resource-constrained environments.


\bibliographystyle{IEEEtran}
\bibliography{references}

\end{document}